\documentclass[onecolumn, preprintnumbers, nofootinbib, aps]{revtex4}
\usepackage{pstricks,graphicx,epsfig,color,amssymb,amsmath,amscd}
\usepackage{latexsym}

\newcommand{\be}{\begin{eqnarray}}
\newcommand{\ee}{\end{eqnarray}}
\newcommand{\rar}{\rightarrow}
\newcommand{\Rar}{\Rightarrow}

\begin{document}

\title{Axion braneworld cosmology}

\author{Cosimo~Bambi$^{\rm 1}$}
\author{Masahiro~Kawasaki$^{\rm 1, 2}$}
\author{Federico~R.~Urban$^{\rm 3, 4}$}

\affiliation{$^{\rm 1}$IPMU, The University of Tokyo, Kashiwa, Chiba 277-8568, Japan \\
$^{\rm 2}$ICRR, The University of Tokyo, Kashiwa, Chiba 277-8582, Japan\\
$^{\rm 3}$UBC, Department of Physics and Astronomy, Vancouver, B.C. V6T 1Z1, Canada\\
$^{\rm 4}$KITPC, Chinese Academy of Sciences, Beijing, 100190, China}

\date{\today}

\preprint{IPMU09-0033}

\begin{abstract}
We study axion cosmology in a 5D Universe, in the case of flat and warped extra dimension.  The comparison between theoretical predictions and observations constrains the 5D axion decay constant and the 5D Planck mass, which has to be taken into account in building 5D axion models.  The framework developed in this paper can be readily applied to other bulk fields in brane universes.
\end{abstract}

\maketitle

\section{Introduction}

One of the open issues of the Standard Model of particle physics is the so called ``strong CP problem''~\cite{Kim:1986ax,Cheng:1987gp,Peccei:2006as,Raffelt:2006cw}. Because of the non-triviality of the QCD vacuum, the effective lagrangian must include the CP violating term
\be
\theta \, \frac{g^2}{32 \pi^2} \, G^2 \, .
\ee
Here $g$ is the QCD gauge coupling constant, $G^2 = G_{\mu\nu}^a \tilde{G}_a^{\mu\nu}$, where $G_{\mu\nu}^a$ and $\tilde{G}_{\mu\nu}^a$ are respectively the gluonic field strength and its dual, and $\theta$ is a dimensionless free parameter.  Measurements of the neutron electric dipole moment demand $\theta \lesssim 10^{-9}$, which signals a very weak CP violation in the QCD sector, and immediately poses the question of why $\theta$ can be so unnaturally small. At present, the most appealing solution seems to be the Peccei--Quinn (PQ) mechanism~\cite{Peccei:1977hh,Kim:1979if,Shifman:1979if,Dine:1981rt,Zhitnitsky:1980tq}, where the PQ symmetry, $U(1)_{PQ}$, is introduced and then spontaneously broken at low temperatures. That generates a Goldstone boson, the axion, which is capable of dynamically cancelling the effective $\theta$-term and thus provides an explanation for the smallness of the $\theta$ parameter.

In the standard 4D theory, the axion anomalous coupling to gluons is
\be
\mathcal{L}_{anom} = \frac{g^2}{32 \pi^2} 
\, \frac{a}{f_a} \, G^2 \, ,
\ee
and the relation between the zero temperature axion mass, $m_a$, and the axion decay constant, $f_a$, is
\be\label{rel-f}
m_a = \frac{z^{1/2}}{1+z} \frac{f_\pi \, m_\pi}{f_a}
= 0.6 \cdot 10^{-3} \, 
\left(\frac{10^{10} \; {\rm GeV}}{f_a}\right) 
\; {\rm eV} \, ,
\ee
where $f_\pi = 92$~MeV and $m_\pi = 135$~MeV are respectively the pion decay constant and mass, while $z= m_u/m_d \approx 0.6$ is the up to down quark mass ratio.

The bounds on the axion decay constant depend on the exact axion model, but the allowed range is essentially~\cite{Amsler:2008zzb}
\be\label{bound-f}
10^8 \; {\rm GeV} \lesssim f_a 
\lesssim 10^{12} \; {\rm GeV} \, .
\ee
The lower bound can be deduced from considerations on stellar cooling or on the neutrino burst of the supernova SN1987A, while the upper limit from the requirement that the present axion relic density does not exceed the dark matter energy density. To be more precise, supernov\ae\ forbid only the range $10^5 \; {\rm GeV} \lesssim f_a \lesssim 10^8 \; {\rm GeV}$, but from other considerations (e.g.\ the overproduction of axions in the early Universe by thermal processes) we find $f_a \gtrsim 10^5$~GeV. Using eq.~(\ref{rel-f}), the constraint~(\ref{bound-f}) can be written as
\be
10^{-5} \; {\rm eV} \lesssim m_a \lesssim 10^{-1} 
\; {\rm eV} \, .
\ee

Axions can be also produced non-thermally via the following (misalignment) mechanism~\cite{Sikivie:2006ni}.  At high temperatures, the cosmic energy density of the axion field is $\rho_a = m^2_a(T) \, a^2$. Here $m_a(T)$ is the thermal axion mass, which has to be calculated by considering QCD instanton effects at high temperature. For $T \approx 1$~GeV, one finds
\be
m_a(T) = 0.1 \, m_a \, 
\left(\frac{\Lambda_{QCD}}{T}\right)^{3.7} \, ,
\ee
where $a^2 \sim (\Theta f_a)^2$ is the value of the axion field. As the temperature of the Universe decreases, the axion mass increases and when it becomes of order the Hubble parameter, $H = 1.7 \, \sqrt{g_*} \, T^2/M_{Pl}$, the axion field starts oscillating and converts its energy into axions. This happens at the temperature $T_m$, given by $m_a(T_m) \approx 3H(T_m)$. The axion number density to entropy density ratio at $T_m$ is
\be\label{number-density}
\frac{n_a}{s} = \frac{m_a(T_m) \, \Theta_{in}^2 \, f_a^2}{s} \, ,
\ee
where $s = (2\pi^2/45) \, g_* \, T_m^3$ ($g_*$ counts the relativistic degrees of freedom at a given $T$). The present value of $\Omega_a = \rho_a/\rho_c$ can be obtained by multiplying eq.~(\ref{number-density}) by the zero temperature axion mass and the present entropy density $s_0 = 2970$~cm$^{-3}$ and by dividing over the critical energy density $\rho_c = 2 \cdot 10^{-29} \, h^2$~g~cm$^{-3}$. The result is
\be
\Omega_a h^2 \approx 0.6 \, 
\Theta_{in}^2 \,
\left(\frac{f_a}{10^{12} \; {\rm GeV}}\right)^{1.18} \, ,
\ee
and implies $f_a \lesssim 10^{12}$~GeV for $\Theta_{in} \sim 1$ ($\Theta_{in}$ is expressed in radians). Actually $\Theta_{in}$ may also be much smaller, say at the level of $10^{-3}$, in order to have $f_a$ close to the GUT scale. However, if the PQ phase transition occurred before inflation, $\Theta_{in}$ is limited by the bound from isocurvature perturbation on the CMB~\cite{Beltran:2006sq,Kawasaki:2007mb,Hertzberg:2008wr}.  But the bound depends on the field value of PQ scalar during inflation which may be different from $f_a$~\cite{Linde:1991km}. Thus the isocurvature constraint is model-dependent and we do not consider this constraint in the present paper.

The scope of this article is to study the cosmology of a 5D axion field in a 5D universe, in which our Universe is thought of an embedded 4D hypersurface.  In particular, we will focus our attention on 5D braneworld models, which are the most studied and discussed in the literature and which represent the simplest extension to our 4D Universe. In the next section, we review the basic ingredients necessary for our analysis, that is, basic facts of 5D axion models and the master equations governing the evolution of the 5D universe. In section~\ref{production}, we study the most important axion production mechanisms (thermal inelastic scattering, time varying metric, coherent oscillation of the axion field) and show for what values of the free parameters of the theory each process turns out to be the most efficient. In section~\ref{bounds}, we present the constraints we can deduce from astrophysical and cosmological arguments in our picture. Summary and conclusions are reported in section~\ref{conclusions}.

{\it Conventions}: Throughout the paper we use natural units $\hbar = c = k_B = 1$. As for the notation for the different gravity scales, $M_{Pl} = 1.2 \cdot 10^{16}$~TeV is the 4D Planck mass, $M_4 = 2.4 \cdot 10^{15}$~TeV the reduced 4D Planck mass and $M_5$ the reduced 5D Planck mass.

\section{5D axion models \label{models}}

In a 5D spacetime, where the Standard Model (SM) particles are confined to the 4D brane (see~\cite{Rubakov:2001kp}), while gravitons and axions live in the bulk, one expects a much richer phenomenology compared to the 4D world~\cite{Dienes:1999gw,Flacke:2006re}, mainly due to:

\begin{itemize}
\item Now the ``fundamental PQ scale'' is the 5D symmetry breaking scale $f_5$. The effective 4D scale, $f_4$, depends on the former and on the shape and the size of the extra dimension;
\item From our low energy 4D viewpoint, a bulk field is seen as discrete or continuous tower of Kaluza-Klein (KK) states;
\item The mass of the zero mode axion may not depend on the PQ scale in the usual way $m_a \sim f_\pi m_\pi/f_a$. To be more precise, in the toy-models of refs.~\cite{Dienes:1999gw,Flacke:2006re}, one finds that the mass of the 0 mode is roughly $\min(1/R,m_a)$, where $R$ is the size of the extra dimension;
\item If the highest temperature at which the Universe attained thermal equilibrium is higher than the so-called $T_t$ (see below), the Universe experiences a period of braneworld expansion~\cite{Cline:1999ts,Binetruy:1999hy,Shiromizu:1999wj}.  This, for example, could imply an efficient gravitational production of axions.  Notice that, in fact, there need not be equilibrium thermal plasma for a non-standard expansion epoch to occur; however, for the phenomenology we are studying here, this is the most relevant case.
\end{itemize}

Some of these considerations apply, of course, to any bulk field, and can be easily extended to other 5D models.

\subsection{Flat extra dimension}

A 5D axion in a flat compact fifth dimension (a circle) is discussed in ref.~\cite{Dienes:1999gw} (note that here we adopt a slightly different notation). The 5D axion action is
\be
\mathcal{S}_{axion} = \int d^4x \, dy
\left[\frac{1}{2}\left(\partial a\right)^2
+ \frac{g^2}{32 \pi^2} \, \frac{a}{f_5^{3/2}} 
\, G^2 \, \delta(y)\right] \, ,
\ee
where $a$ is the 5D axion field, whose energy dimension is 3/2. Expanding $a$ into Fourier modes
\be
a(x,y) = \frac{1}{\sqrt{2 \pi R}} \sum_n a_n(x) \, b_n(y) 
\ee
and integrating over the extra dimension, we get the effective 4D lagrangian. The effective 4D PQ scale is $f_4 = (2 \pi R f_5)^{1/2} f_5$ and, for $R f_5 \gg 1$, it can be much higher than the 5D decay constant $f_5$ (the picture is like that of the 4D and 5D Planck masses, which are related to each other by $M_4 = (2 \pi R M_5)^{1/2} M_5$).

Physical phenomena can be easily described and interpreted using a 4D or a 5D point of view. In the first case, we have KK axions with coupling constant $f_4$ and mass splitting $1/R$ (this is valid for high KK modes, and is only an approximation for lower states, see below): the axion production rate at the energy $E$ is given by the the production rate for a single state, $\sim E^3/f_4^2$, times the number of states with mass smaller than $E$, i.e.\ $RE$ if $RE \gg 1$, $\Gamma \sim E^4 \, R / f_4^2$.  Adopting the 5D point of view, there is just one axion field, but the axion coupling is $1/f_5^{3/2}$. The rate is $\Gamma \sim E^4 / f_5^3$, and the two descriptions are equivalent, since $f_4^2 \sim R f_5^3$.

Let's write down the few equations that will enable us to switch from one set of parameters to another (in TeV units).
\be
1/R &\simeq& 1.1 \cdot 10^{-30} M_5^3 \, ;\\
f_4 &\simeq& 2.4 \cdot 10^{15} (f_5 / M_5)^{3/2} \, ;\\
m_a (eV) &\simeq& 2.5 \cdot 10^{-12} (M_5 / f_5)^{3/2} \, ;\\
m_0 &=& {\rm min}\{1/R, m_a\} \, .
\ee

Finally, in general the mass of each KK mode is given approximately by $m_n = m_0 + n/R$, where $R^{-1} = 2\pi M_5^3 / M_4^2$ is the inverse of the size of one compact extra dimension. This expression of course is not universally true, since an effective potential for the 5D field may appear upon dimensional reduction, and therefore contribute to the effective mass of the KK mode. This is indeed what happens in the extra dimensional axion models at hand, but nevertheless, it is not difficult to convince oneself that at high $n$ this parametrisation provides a very good approximation.  This applies to the warped case as well.

\subsection{Warped extra dimension}
The case of one warped extra dimension is discussed in ref.~\cite{Flacke:2006re}. However, the picture is quite similar to the flat case and likely there are no relevant differences.

If the extra dimension is warped, the structure of the KK tower turns out to be subtly different. In the case of two branes, typically one finds that the mass of the KK mode $n$ is
\be
m_n = m_0 + k x_n e^{-\pi R k} \, ,
\ee
where $k = (1 - \exp (-2\pi k R)) M_5^3 / M_4^2$ is the $AdS_5$ curvature, $x_n$ is the $n$-th root of the first order Bessel function $J_1$ and $\pi R$ is the size of the orbifold. The mass splitting is $\Delta m \simeq 3 k \exp(- \pi R k)$, and for later convenience we define the function $F(kR) \equiv (1 - \exp (-2\pi k R)) / \exp (\pi k R)$.

Thus, some care should be taken in working out the formulas for the abundances, as the expressions for the mass gaps are different.  However, as it will be shown below, in the end the picture is almost unchanged.

\subsection{Brane cosmology}
As it has been noticed at the beginning of this section, one of the main differences between 4D and 5D cosmologies lies on the different relation linking time and temperature.  Indeed, if the SM lived on a four dimensional Friedman--Robertson--Walker hypersurface (the brane), embedded in an extra dimensional spacetime, the early universe would admit an epoch of non-standard expansion~\cite{Shiromizu:1999wj}.  Several such models have been built in the last few years, see~\cite{ArkaniHamed:1998rs,Randall:1999ee,Randall:1999vf} for some of the original proposals.  These models show a peculiar feature when their cosmology is investigated.  Indeed, a general feature of the 5D embedding is that on each 4D brane the Friedman equation has the form~\cite{Cline:1999ts,Binetruy:1999hy,Shiromizu:1999wj}:
\be\label{modFried}
H^2 = \frac{\rho}{3M_4^2} \left( 1 + \frac{\rho}{2\lambda} \right)\, ,
\ee
where $\lambda$ is the tension of the brane, which is related to the five dimensional Planck mass as $\lambda=6M_5^6/M_4^2$. This equation says that at high energy densities the expansion of the universe was much faster than at later times, and went as $T^4$ instead of $T^2$, together with the unknown parameter $M_5$: the smaller $M_5$, the faster the expansion.  Here we have ignored further terms in the effective 4D equation, but they are required to be very small by observation, see for instance~\cite{Maartens:2003tw}.

At this point it is convenient to define a ``transition'' temperature $T_t$ from standard cosmology to brane one, which can be extracted from $\rho=2\lambda$:
\be\label{trT}
T_t^2 = 
\left(\frac{360}{\pi^2 \, g_*}\right)^{1/2}\, \frac{M_5^3}{M_4}\, ,
\ee
where $g_* = g_*(T)$ counts the relativistic degrees of freedom at a given temperature $T$. If the dominant component of the universe is not radiation, then this ``temperature'' approximately means the fourth root of the energy density, and parametrises the epoch at which the transition occurs.

One basic requirement is that the Universe expands as $H^2\propto\rho$ at the BBN, which means $T_t \gtrsim 1$~MeV, or $M_5 \gtrsim 10$~TeV.

\section{Production Mechanisms \label{production}}

\subsection{Thermal Production}
In this section we will be analysing the conditions under which a thermal population of (KK) axions is born in the early Universe.  The results obtained here are not only relevant for the thermally generated abundances, but also in dealing with other non-thermal production mechanisms (gravitational particle production, misplacement mechanism, etc), as non-thermal abundances generated before the epoch at which axions thermalise will be phagocytosed by the thermal bath.

The most important process which thermalises axions in the early Universe (assuming inflation takes place at higher scales) is their interaction with QCD matter. The rate of these processes (see~\cite{Masso:2002np}) can be estimated as
\be\label{rate}
\Gamma \simeq 7 \cdot 10^{-6} T^3 / f_4^2 \simeq 1.2 \cdot 10^{-36} T^3 M_5^3 / f_5^3 \, ,
\ee
and the expansion rates of the Universe are given by (in TeV units)
\be\label{exp}
H_{std} \simeq 1.4 \cdot 10^{-16} \sqrt g_* T^2 \, ,\\
H_{bc} \simeq 6 \cdot 10^{-2} g_* T^4 / M_5^3 \, .
\ee
It is straightforward then to realise that thermal equilibrium took place when $\Gamma \gtrsim H$, which implies that 
\be
T^{eq}_{std} \gtrsim T^D_{std} \simeq 1.1 \cdot 10^{20} \sqrt g_* f_5^3 / M_5^3 \label{eqTstd} \, ,\\
T^{eq}_{bc} \lesssim T^D_{bc} \simeq 2 \cdot 10^{-35} g_*^{-1} M_5^6 / f_5^3 \label{eqTbc} \, ,
\ee
where $T^{D}_{std}$ and $T^{D}_{bc}$ are the decoupling temperatures, respectively in the standard and braneworld expansion period. The transition temperature between the two different regimes (from $H^2 \sim \rho^2$ at high energies to $H^2 \sim \rho$ at late times) happens at
\be\label{Tt}
T_t \simeq 5 \cdot 10^{-8} g_*^{-1/4} M_5^{3/2} \, .
\ee
In order for the two equations~(\ref{eqTstd}) and (\ref{eqTbc}) to make sense, the following condition must be satisfied:
\be\label{f5consistency1}
T^D_{bc} \gtrsim T_t \gtrsim T^D_{std} \quad\Rar\quad f_5 \lesssim 8 \cdot 10^{-10} g_*^{-1/4} M_5^{3/2} \, , 
\ee
or the band of temperature for which thermal equilibrium is realised would shrink to nothing. Notice further that if the reheating temperature or the 5D PQ breaking scale were smaller than the upper limit $T^D_{bc}$, then they would be the actual upper (in temperature) limit at which thermal equilibrium ceases to be realised. Of course, in the case the reheating temperature is even smaller than $T_{std}^D$, axions were never in equilibrium after inflation.

Now, for each mode in thermal equilibrium which decouples when still relativistic, we have:
\be\label{Ynth}
Y_n = \frac{n_n}{s} = 0.278 / g_* \, .
\ee
Here, as previously, $g_*$ is the number of relativistic degrees of freedom in equilibrium with the thermal plasma. In our case (formula~(\ref{Ynth})) this is taken at the decoupling temperature. Moreover, if there are many KK axion states in thermal equilibrium, this number is given by $g_{SM} + n_{KK}$ with $n_{KK}$ accounting only for the KK states which are relativistic and in equilibrium, and $g_{SM} = 106.75$ for the SM, or about twice as much for its supersymmetric version.

In order to render the analysis traceable and clear, it is most useful at this point to specify a working value for $f_4$.  With a little foresight, we pick the value $f_4 = 10^7$ TeV: we will comment on the dependence of our results on this choice every time it will turn out to be relevant.

By fixing a value for $f_4$ we automatically tie $f_5$ to $M_5$ as
\be\label{f5spec}
f_5 \simeq 3 \cdot 10^{-6} M_5 \, .
\ee

{\bf\emph{Flat fifth dimension --}} 
Since from this point on we will need the explicit expression for the mass gap, it is sensible to split the analysis in two, flat and warped 5D axion models.

First of all, let us look now at the behaviour of the effective number of light degrees of freedom with the 5D Planck mass.  As we are ultimately interested in this number at decoupling, we write the number of relativistic axion modes at decoupling as $n_{KK}^D = T^D_{std} R$.  There is therefore a critical value for $M_5$ at which the equilibrium KK axions overwhelm the SM degrees of freedom;  this happens around $M_5 \approx 5.6 \cdot 10^{10}$~TeV.  This special value of $M_5$ marks the point at which, at decoupling, the number of equilibrium KK axions reaches (or drops below) the number of SM degrees of freedom.  Consequently, the decoupling temperature $T^D_{std}$ becomes
\be
T^D_{std} \simeq 3 \cdot 10^4 \, \textrm{TeV} \qquad &\textrm{when}& \qquad M_5 \gtrsim 6 \cdot 10^{10} \, \textrm{TeV} \, , \label{m5eqStd}\\
T^D_{std} \simeq 4 \cdot 10^{36} M_5^{-3} \, \textrm{TeV} \qquad &\textrm{when}& \qquad 1.2 \cdot 10^{10} \mathrm{TeV} \lesssim M_5 \lesssim 6 \cdot 10^{10} \, \textrm{TeV} \, , \label{m5eqBc}
\ee
where the first line applies to $g_* \simeq g_{SM}$ and the second case is instead $g_* \simeq T^D_{std} R$.  Hence, the number of equilibrium KK states that decouple at $T^D_{std}$ is almost always smaller than the SM ones, unless $1.2 \cdot 10^{10} \, \textrm{TeV} \lesssim M_5 \lesssim 6 \cdot 10^{10} \, \textrm{TeV}$.

The picture thus is as follows. In the early Universe, even before $T_t$ we have a population of thermal axions because, as long as~(\ref{f5consistency1}) holds, the window for thermalised KK axions is open, and goes from $T^D_{bc}$ down to $T^D_{std}$, crossing $T_t$.  The number of states that are still relativistic and in thermal equilibrium at the lowest decoupling temperature depends on the value of $M_5$ as shown in~(\ref{m5eqStd}) and (\ref{m5eqBc}).  Since we have assumed that the cross section is independent on the mass of the KK state, then all these states will come out of equilibrium at the same time when $T = T^D_{std}$, and all of them will be by definition relativistic then.  Of course also axions of higher mass were in thermal equilibrium before the collective decoupling, but they were non relativistic and their abundances are therefore irrelevant due to Boltzmann exponential suppression.  Once the axions decouple, their yield variable remains pretty much the same until today, unless some modes decay (or other production mechanisms) turn on.

Given these results and this picture in mind, we can rewrite the yield variable at decoupling as
\be\label{yield}
Y_n(t_D) \simeq 0.278 / g_* \simeq
\left\{
\begin{array}{ll}
3 \cdot 10^{-3} & \textrm{when} \qquad M_5 \gtrsim 6 \cdot 10^{10} \, \textrm{TeV} \\
&\\
9 \cdot 10^{-68} \, M_5^6 \, & \textrm{when} \qquad 1.2 \cdot 10^{10} \, \textrm{TeV} \lesssim M_5 \lesssim 6 \cdot 10^{10} \, \textrm{TeV}
\end{array} \right. \, ,
\ee
which is valid per each d.o.f. up to $n_{KK}$.

Probably the easiest way to analyse the impact of KK towers in late time cosmology is to split the tower into mass bands according to the constraints one is going to look at.  This is so because different constraints corresponds to different decay time ranges, which in turn, once the zeroth mode decay width has been given, depend almost solely on the masses.  Once the interesting lower and upper limits have been identified, one is in a position to scrutiny the number of modes in that given range, and investigate their total impact on a particular observable. Moreover, it is reasonable to expect that only large number of states for each band could provide significantly different constraints compared to an ordinary 4D particle with mass within the given band. This last consideration, combined with the fact that the zero mode is supposed to be light, implies that the highest KK mode $N$ in the band will satisfy $N \gg N_0$, with $N_0$ the lightest mode in the same band.  This scheme was endorsed in~\cite{Okada:2004mh,Bambi:2007vk} to investigate thermal production of 5D gravitinos, and in~\cite{Bambi:2008ch} for gravitationally produced particles (see also below).

By looking at~(\ref{yield}) once can readily evince that single modes, in order to be safe, demand:
\begin{itemize}
	\item Small mass, therefore a small $M_5$
	\item A high $T^D_{std}$ which implies more dilution afterwards. This wants a high $f_5$ (not displayed in our formulas), although there is a threshold beyond which thermal equilibrium is never realised; it also requires a small $M_5$, again.
	\item Much dilution, that is, the largest $n_{KK}$ available, which again supports small $M_5$ and high $f_5$.
\end{itemize}

The situation changes drastically when a summation is involved. Indeed in this case the solution to overproduction of axions is a widening of the mass gap between different states, which implies on the one hand less dilution (which is bad), but on the other hand precludes an enormous number of states to be summed over and contribute to the total energy density today. This can be put in formulas as
\be\label{Ytot}
\frac{\rho_t}{s} = \sum_{n=0}^{N} m_n Y_n = \left\{
\begin{array}{ll}
1.4 \cdot 10^{-33} \, N^2 \, M_5^3 \, \textrm{TeV} & \textrm{when} \qquad M_5 \gtrsim 6 \cdot 10^{10} \, \textrm{TeV} \\
&\\
5 \cdot 10^{-98} \, N^2 \, M_5^9 \, \textrm{TeV} & \textrm{when} \qquad 1.2 \cdot 10^{10} \, \textrm{TeV} \lesssim M_5 \lesssim 6 \cdot 10^{10} \, \textrm{TeV}
\end{array} \right. \, ,
\ee
which (remember that $N \propto M_5^{-3}$) shows the inverse dependence on $M_5$, thereby imposing a lower limit on $M_5$.  Notice that this is not the case when the number of equilibrium states is dominated by KK axions, but it will be true for the other production mechanisms.

{\bf\emph{Warped fifth dimension --}}
If the extra dimension is warped the mass gap is much different than in a flat bulk, difference which is traceable by following the effects of the function $F(kR)$ in the previous formulas.

In order to have an idea of what happens in this case, let's work out the equilibrium conditions (analogously to eq.~(\ref{yield})).  Firstly, it is necessary to know which degrees of freedom dominate at decoupling.  In doing so, we separate the problem in two, by noticing that while the highest value $F$ can take (that is, around 0.4) would leave the coefficients unchanged, for much smaller $F$'s (equivalent to more pronounced warping) the situation is dramatically different.

Indeed, while in the first case there are only tiny numerical differences, if the warping factor grows there are more and more KK axions in thermal equilibrium (for fixed $M_5$) and if we take $kR=11$ the KK equilibrium d.o.f. always dominate the SM ones. In this second case, thermal equilibrium is realised as long as $M_5 \gtrsim 8.1 \cdot 10^{13} \, \textrm{TeV}$, and the predicted thermal abundance is
\be
Y_n(t_D) \simeq 1.9 \cdot 10^{-98} M_5^6 \, .
\ee
This, as long as $M_5\gtrsim 10^{13}$~TeV, would be a disastrous amount of KK particles ($\gtrsim 10^{-14}$ per each mode), but they luckily decay just before the onset of the BBN and, therefore, are reasonably harmless.

\subsection{Gravitational production}
Gravitational particle production in an ordinarily expanding Friedman-Robertson-Walker (FRW) Universe dominated by dust-like or radiative matter is known to be a very poorly efficient particle creation mechanism~\cite{Birrell:1982ix,Mamaev:1976zb}.  However, as it has been recently pointed out in ref.~\cite{Bambi:2007nz}, if the four dimensional FRW Universe were to be embedded into a higher dimensional spacetime, then even dust-like or radiation dominated FRW Universes would be able to inject gravitationally produced particles into the plasma in a sizeable way. This is easily understood as a consequence of the possibility that the actual scale of - 5D - gravitational interaction be tuned to much lower values, thereby drastically enhancing the coupling with matter.

In what follows we simply borrow the results of \cite{Bambi:2008ch,Bambi:2007nz}, referring the interested reader to these papers for the details.

{\bf\emph{Flat fifth dimension --}} Having obtained explicit formulas for the KK masses, one can readily write down the gravitationally produced abundance associated to each KK mode. In the limit for which the zero mode is light compared to $n/R$, the yield variable turns out to be
\be
Y_n \simeq 1.2 \cdot 10^{-71} n^{9/4} \, M_5^{9/2} \, . \label{yNflat}
\ee

In order for all this to be consistent, we need to ensure that
\be
T_f \gtrsim T_t &\Rar& n \gtrsim 0.3 \, ,
\ee
What this means (recall it is an order of magnitude estimate) is that the result~(\ref{yNflat}) does not hold for the first handful of modes, as the quantities $\Omega_n$ and $Y_n$ have been deduced assuming $H = \rho/6 M_5^3$, while for small $n$ we would find $T_f < T_t$. For these light modes, the gravitational production stops at the transition temperature $T_t$ or, if very light, during the standard expansion, thereby rendering their abundances negligible.

With the single mode contributions at hand, it is now straightforward to obtain the overall yield variable for each band by just summing over the modes (up to the highest mode $N$) and then discarding the (smaller) contribution of the lowest state, the result of which being
\be
\frac{\rho_g}{s} \equiv \sum_n m_n Y_n \simeq 3 \cdot 10^{-102} N^{17/4} \, M_5^{15/2} \, . \label{yTOTflat}
\ee

{\bf\emph{Warped fifth dimension --}} The main difference arising in this case is the different expression for the KK masses, as the Friedman equation is basically the same of the case with a flat extra dimension and we can still use the modified Friedman equation. It is thus fairly straightforward to repeat the steps undertaken for the flat case, with result
\be
Y_n \simeq 2 \cdot 10^{-72} \, n^{9/4} F^{9/4} M_5^{9/2} \label{yNwarped} \, ,
\ee
in place of eq.~(\ref{yNflat}).

Similarly to the case of thermal axions, the scenario is more or less unchanged in this picture, unless one employs very small values for $F(kR)$. Indeed this function is always smaller than $\sim 0.4$, for which value the abundances and consistency constraints turn out to be very similar to those previously obtained for a flat extra dimension. If one instead chooses to work with much smaller $F(kR)$, such as for $kR = 11$, then the mass splitting becomes extremely tiny, unless the five dimensional Planck mass is pushed all the way up to the 4D $M_4$. Therefore, the mass gap drops by a factor of $2/F$, and the abundances of produced particles increase by a huge factor $1/F$. The compensation in $M_5$ is proportional to $F^{-4/21}$ which, again for $kR = 11$, is around 700, although in highly warped models it is customary to safely (as far as this mechanism is concerned) take $M_5 \simeq M_4$.

Along these lines one can now transfer to the warped solution the results which will be presented below for the flat scenario.

\subsection{The misalignment mechanism}

{\bf\emph{Flat fifth dimension --}} In the standard scenario, (often) the most efficient production mechanism of axions in the early Universe is associated to field oscillations. Such a mechanism works because massless (or light, that is $H \gg m$) scalars are unstable in $dS$ spacetime and during inflation can acquire a large vacuum expectation value (vev). Then, when the expansion rate of the Universe becomes smaller than $m$, the field goes to the minimum of its potential and converts its energy into particles. In order to know how many modes have acquired a large vev, we should know the value of the Hubble parameter during inflation, $H_I$. The number of modes would be $N \approx RH_I$.

Other important inputs are the temperature dependent axion masses (temperature corrections are likely significant for the zero and the lightest KK modes, while are more probably irrelevant for the other ones) and the kind of expansion (standard or brane regime?). In other words, the estimate of the axion abundance depends on several parameters, which we will try to track in our analysis.

Let us assume that the PQ symmetry is already broken after inflation. In this case, the axion field is homogeneous over large distances and we have to consider only the zero momentum mode\footnote{If the PQ phase transition is after inflation, that is not true and we have to consider non-zero momentum modes as well, whose contribution to the axion cosmological energy density would be similar to the one of the zero momentum mode. So, neglecting such a possibility we find at most a more conservative bound.}.

The abundance of KK modes can be evaluated in a very similar way to that described for the zero mode, but now using the KK mass $m_n = n/R = 2\pi n M_5^3/M_4^2$, and assuming that the thermal mass is not relevant (which will be the case for heavy KK states, which in turn are the states we are mostly dealing with).  Since we are interested only in the heavy KK modes, we will not go into the details of the computation of the thermal mass. For $T_m^n>T_t$, the $n$-th KK mode oscillates during the braneworld period and we get
\be
T_m^n &=& 1.6 \cdot 10^{-8} \, n^{1/4} \, M_5^{3/2} \; \mathrm{TeV} , \\
Y_n &=& 6 \cdot 10^{6} n^{1/4} \Theta_{in}^2 \, M_5^{-3/2} \, . \label{n-mode-mis-1}
\ee

For $T_m^n < T_t$, the $n$-th mode oscillates when the Universe expands in the standard way and we find
\be
T_m^n &=& 1.6 \cdot 10^{-8} \, n^{1/2} \, M_5^{3/2} \; \mathrm{TeV} , \\
Y_n &=& 6 \cdot 10^{6} n^{-1/2} \Theta_{in}^2 \, M_5^{-3/2} \, . \label{n-mode-mis-2}
\ee
The total axion abundance can be easily deduced by summing over all the KK modes which acquired a large vev during the inflation period. That is determined by the exact model of braneworld inflation. A more accurate estimate of axion production from coherent oscillation would require a numerical integration of the field equation in the expanding background. Nevertheless, such an effort would be likely useless, because the most important parameter is the initial misalignment angle $\Theta_{in}$ which is unknown.

Notice also that the temperature $T_m$ is always (for our choice of $f_4$) above the transition temperature, except for the first few KK modes.  Therefore, in the discussion of the constraints, we will employ only the abundance calculated from (\ref{n-mode-mis-1}).

A comment is in order here, concerning the analysis of relic axion oscillations presented in~\cite{Dienes:1999gw}.  There a mechanism for enhancing the rate of dissipation of the total (zero mode plus KK states) energy density stored in the axion condensate is exploited for rendering the 5D axion model possibly even safer than the standard 4D scenario.

This mechanisms relies on one fundamental assumption, that is that all KK modes sits at the minimum of the potential ($\Theta_{in} = 0$) when the zero mode begins its oscillations.  In this case the zero mode triggers these higher KK modes into oscillations, and therefore transfers some of its energy density into them.  As is well known, heavier axions during the subsequent evolution of the Universe lose energy more efficiently than light ones, which is why axions produced by the misalignment mechanism put a bound on the invisibility of the axion (that is, require a minimum mass).  Hence, the net result is that KK modes help the dissipation of the total energy density in this model.

However, since heavy KK modes are expected to decay, the dissipation happens through the first few KK states.  But these states are also the states that are likely to have some initial displacement from the bottom of their potentials (which appears even before the QCD transition due to the mass matrix between KK states).  This initial displacement is not accounted for in~\cite{Dienes:1999gw}, and we therefore expect that the benefits of this mechanism will not hold in a more realistic case.

One last brief note before moving to the warped scenario.  In the analysis of~\cite{Dienes:1999gw} the effects of the brane regime were not included; indeed if one looks at the values of $f_4$ and $M_5$ needed to have an enhanced dissipation rate, one finds that, for $f_4 \approx 10^7$ TeV, the 5D Planck mass (assuming only on extra dimension) needs to be below 500 TeV, but we know that then the zero mode will start oscillating before $T_t$, and the equation of motion will therefore be different (due to the different dependence of $H$ upon time).

We can now write down the total abundance, keeping $\Theta_{in}$ as an unspecified parameter, as
\be
\frac{\rho_m}{s} \simeq 3 \cdot 10^{-24} N^{9/4} \Theta_{in}^2 M_5^{3/2} \, .
\ee

{\bf\emph{Warped fifth dimension --}} 
Like for the other production mechanisms, here we have just to replace the KK mass of the flat model, $m_n = 2\pi n M_5^3/M_4^2$, with the one of the warped model, $m_n = 3 F(kR) n M_5^3/M_4^2$.  The same considerations presented for the flat case apply to this scenario as well, and will change the results of ref.~\cite{Flacke:2006re}.

\section{Constraints \label{bounds}}
In this section we will be analysing the bounds coming from different cosmological arguments, such as the density of Dark Matter particles today, BBN constraints, CMB distortion, X--rays, effective number of neutrinos, and recombination.  In order to do that, we need to know the lifetime of the KK axions. For masses larger than about 100~MeV, the main decay mode is usually into hadrons and the lifetime is
\be\label{tau-had}
\tau_{\mathrm{had}} \sim \left(\frac{\alpha_s^2}{256 \, \pi^3} \, 
\frac{m_a^3}{f_4^2}\right)^{-1}
= 5 \cdot 10^{26} \, \alpha_s^{-2} \, 
\left(\frac{1 \; {\rm eV}}{m_a}\right)^3 
\left(\frac{f_4}{10^7 \; {\rm TeV}}\right)^2 \; {\rm s} \, .
\ee
For smaller masses, the decay into hadrons is kinematically forbidden, and the main channel is $a \rar 2\gamma$. The lifetime is given by eq.~(\ref{tau-had}) with $\alpha_s$ replaced by $\alpha_{em}$:
\be
\tau_\gamma \sim 1.0 \cdot 10^{31} \, 
\left(\frac{1 \; {\rm eV}}{m_a}\right)^3 
\left(\frac{f_4}{10^7 \; {\rm TeV}}\right)^2 \; {\rm s} \, .
\ee
In the end of this section we will briefly reconsider the supernova bound as well.  All results refer to the flat extra dimension case, but it is straightforward to export the results to the corresponding warped scenario, thanks to the explicit formulas for the abundances computed in the previous section.

Before embarking on to the discussion of all these limits, a general remark on thermal abundances is in order.  We have seen that thermal equilibrium for the axion (at $f_4 = 10^7$~TeV) occurs only if $M_5 \gtrsim 10^{10}$~TeV or so.  However, this implies automatically that the first KK mode will have a mass of order TeV, which implies that it will decay much faster than 1s, and therefore provide no constraint at all.  This will not in general be true for different values of $f_4$, and more in general for other bulk fields, but in this specific case thermal abundances of KK states are automatically made safe.

A description of each way we can constrain the abundance of KK axions follows, together with the rough estimates for the limits on the 5D Planck mass.  A summary table with the more precise limits is presented at the end.

\subsection{Cold relics}
Axions whose lifetime exceeds that of the Universe $t_0 \approx 10^{17} \div 10^{18}$~s, will contribute to the energy density of invisible matter today, which is constrained by observations to be not in excess of $\Omega_a h^2 \approx 0.12$.  This can be translated as
\be
\frac{\rho_a}{s} \lesssim 5 \cdot 10^{-13}\mathrm{TeV} \left( \frac{\Omega_{\mathrm{DM}} h^2}{0.12} \right) \, .
\ee

The number of KK states which have survived long enough to be counted in this quantity are all the light ones up to $N \simeq 3 \cdot 10^{22} \, M_5^{-3}$, and their contributions due to gravitational and misalignment abundances are
\be
\frac{\rho_g}{s} &\simeq& 1.5 \cdot 10^{-6} M_5^{-21/4} \, ,\\
\frac{\rho_m}{s} &\simeq& 1.3 \cdot 10^{27} \Theta_{in}^2 M_5^{-21/4} \, ,
\ee
which in turn means that the 5D Planck mass has to be bigger than about 17 TeV and $3 \cdot 10^7 \, \Theta_{in}^{8/21}$~TeV, respectively.

Notice that these constraints, as all those that follow, are valid only if there effectively is a densely populated tower of states within the mass range considered, fact which can be easily checked by comparing the highest mass for each band with the minimum mass gap required by the corresponding constraint.  As an example (which can be reiterate in the following discussion), in the case just worked out the highest mass involved is of order 40~keV, and the minimum mass gaps are around $10^{-15}$~eV and $40\,\Theta_{in}^{24/21}$~keV for gravitational abundances and misalignment ones, respectively.  Therefore, while the result obtained from gravitational production is surely valid, the one referring to the misalignment mechanism is good only for small angles $\Theta_{in}$.  This is a general result which is found to hold in the forthcoming sections as well.

\subsection{BBN}
If a particle lives long enough to witness the synthetisation of the light elements (Big Bang Nucleosynthesis, or BBN), but decays during this process or shortly thereafter, the entropy density injected into the equilibrium plasma can potentially alter the final abundances of these elements, thereby causing conflicts with the good matching between theoretical predictions and observations.  Limits on the amount of energy density to entropy ratio which can be safely transferred to the plasma are usually among the strongest ones for particles whose lifetime is in the range of interest $\tau~1$~s to $\tau~10^{13}$~s.  In working out the constraints on the KK axion abundances we closely follow the schematisation of ref.~\cite{Kawasaki:2007mk}, and we denote $B_\gamma$ and $B_{\mathrm{had}}$ the branching ratios into photons and hadrons, respectively.

The four possibilities we consider are: 1) $B_\gamma=1$ with $10^4\,\mathrm{s}\lesssim\tau\lesssim10^7\,\mathrm{s}$; 2) $B_\gamma=1$ with $10^7\,\mathrm{s}\lesssim\tau\lesssim10^{11}\,\mathrm{s}$; 3) $B_{\mathrm{had}}=1$ with $1\,\mathrm{s}\lesssim\tau\lesssim10^4\,\mathrm{s}$, and finally; 4) $B_{\mathrm{had}}=1$ with $10^4\,\mathrm{s}\lesssim\tau\lesssim10^{11}\,\mathrm{s}$.  This last band will not be able to constrain the parameters of the extra dimensions, since the hadronic axions with masses above 100~MeV decay before $10^4$s.  Finally, notice that, although limits can be obtained for decays down to $\tau~10^{13}$~s, the existence of the 4.5~MeV threshold (below which there is not enough energy to dissociate D + $\gamma \rar$ n + p), effectively cuts the lower end of the bands around $10^{11}$~s.

Moreover, for the sake of simplicity, the allowed total $\rho_a / s$ have been flattened over the timescales (or mass ranges) pertaining to each band, which means that our constraints are actually conservative.

The three possibilities are reported in table \ref{table}, where we can see that gravitational particle production limits the lower value of $M_5$ around $10^4$~TeV to $10^5$~TeV, while, although some care should be taken when reading these limits, misalignment axions tend to push $M_5$ up to values for which the first KK mode should have decayed before 1s.  However, if the misalignment angle is small enough (smaller than $10^{-4}$ for example), then the strongest limits are comparable to those coming from gravitational production.

\subsection{CMB distortions}
If the KK states have a lifetime between $10^6$~s and $10^{13}$~s, then their decay may affect the blackbody spectrum of
CMB.  Indeed, such decays would transfer energy to the plasma at the epoch when processes such as the double-Compton scattering, necessary to preserve the blackbody spectrum of the CMB, become inefficient.  The distortion of the spectrum is characterised by the chemical potential $\mu$ at times greater than $10^9$~s when the energy transfer by
the Compton scattering is efficient, and Compton $y$-parameter at later epoch.  Constraints on these quantities can be read off~\cite{Fixsen:1996nj}, and, once translated into upper bounds on the energy density to entropy ratio, become
\be
\frac{\rho_a}{s} \lesssim 3 \cdot 10^{-14} \, \mathrm{TeV} &\qquad\mathrm{for}\qquad& 10^6 \, \mathrm{s} \lesssim \tau \lesssim 10^9 \, \mathrm{s} \, ,\\
\frac{\rho_a}{s} \lesssim 7 \cdot 10^{-16} \, \mathrm{TeV} &\qquad\mathrm{for}\qquad& 10^9 \, \mathrm{s} \lesssim \tau \lesssim 10^{13} \, \mathrm{s} \, .
\ee

The resulting lower limits on the 5D gravity scale $M_5$ are similar to those extracted from BBN: around $10^4$~TeV for the gravitational abundances, and $10^9 \, \Theta_{in}^{8/21}$~TeV for the KK axions produced via the misalignment mechanism.  Notice that, as seen in the summary table below, KK axions that decay around $10^6$s have masses slightly above the threshold for the decay into hadrons, so the limits we are quoting are valid only for $B_{\mathrm{had}}\ll1$.

\subsection{Effective number of neutrinos}
The expansion rate of the Universe depends on its total energy density.
The study of phenomena which are sensitive to the Hubble parameter can 
thus put constraints on the amount of new particles. The best example 
is the BBN, where the primordial production of light nuclei depends on 
the expansion rate when the Universe temperature was in the range
$1 \; {\rm MeV} - 10 \; {\rm keV}$, that is, at the time 
$t \approx 1 - 10^4 \; {\rm s}$ (see e.g.\ ref.~\cite{villante}). The bound that 
can be deduced is often expressed in the literature as the effective 
number of neutrinos and current data demand no more than one extra 
neutrino species: $\Delta N_\nu \lesssim 1$.

The KK axions which are stable or quasi-stable at the BBN are the ones
which can not decay into hadrons and whose mass is not larger than about
100~MeV. If we assume $1/R \ll 100$~MeV, i.e. $M_5 \ll 5 \cdot 10^8$~TeV,
the highest KK mode is $N \sim 10^{26} \, M_5^{-3}$. The total KK axion
energy density to entropy density ratio due to gravitational and 
misalignment productions are respectively
\be
\frac{\rho_g}{s} &\sim& 7 \cdot 10^8 \, M_5^{-21/4} \; ,\\
\frac{\rho_m}{s} &\sim& 7 \cdot 10^{34} \,
\Theta_{in}^2 \, M_5^{-21/4} \; .
\ee
Such a quantity must be smaller than $1.3 g_*^{-1}(T_{prod}) T_{BBN}$, 
where $g_*(T_{prod})$ counts the number of degrees of freedom at the
time of axion production, while $T_{BBN}$ is the temperature at the BBN.
So, BBN demands $M_5 \gtrsim 1.52 \cdot 10^3$~TeV, if we consider the
mechanism of gravitational production, and
$M_5 \gtrsim 1.4 \cdot 10^8$~TeV in the case of coherent oscillations 
of the axion field ($\Theta_{in} = \pi$).

\subsection{X--ray background}
Photons with energy in the range $1 \; {\rm keV} - 1 \; {\rm TeV}$ 
produced at time $t \gtrsim 10^{13}$~s can contribute to the diffuse 
X--ray background. The flux of photons from the two photon decay 
of a particle of mass $m$ and lifetime $\tau$ is found to 
be~\cite{yanagida}
\be
\Phi_{th} (E) &=&
\frac{E}{4\pi} \int_0^{t_0} dt' \, 
\frac{B_\gamma \, n(z)}{\tau (1+z)^3} \,
\frac{dE'}{dE} \, 2 \delta\left(E' - m/2\right) = \nonumber\\
&=& \frac{B_\gamma s_0 Y}{2 \pi \tau H_0}
\, f(m/2E) \, \exp\left[\frac{1}{3 H_0 \tau \Omega_\Lambda^{1/2}}
\, \ln \left( \frac{f(m/2E) \Omega_\Lambda^{1/2} - 1}{f(m/2E) 
\Omega_\Lambda^{1/2} + 1} \frac{\Omega_\Lambda^{1/2} 
+ 1}{\Omega_\Lambda^{1/2} 1 1} \right)\right] \; ,
\ee
where $B_\gamma$ is the branching ratio into $2\gamma$, $n(z)$ is
the particle number density at redshift $z$, $E'$ is the photon
energy at the instant of production, $E$ is the present photon
energy,
\be
f(m/2E) = \left[\Omega_\Lambda + \Omega_m 
\left(\frac{m}{2E}\right)^3\right]^{-1/2}
\ee
and in the last step we assumed that the Universe is flat, 
i.e.\ $\Omega_\Lambda + \Omega_m = 1$. The estimated photon
flux must be smaller than the observed one~\cite{asca, heao1}:
\be
\Phi_{obs} &=& 8 \, 
\left(\frac{E}{1 \; {\rm keV}}\right)^{-0.4} \; 
{\rm cm^{-2} \, s^{-1} \, sr^{-1}} \; , 
\ee
in the energy range $E \approx 0.2 - 25$~keV, and
\be
\Phi_{obs} &=& 6 \cdot 10^{-3} \, 
\left(\frac{E}{1 \; {\rm MeV}}\right)^{-1.6} \; 
{\rm cm^{-2} \, s^{-1} \, sr^{-1}} \; .
\ee
for $E \approx 25 \; {\rm keV} - 4 \; {\rm MeV}$.

The strongest constraint comes from KK axions with lifetime much 
shorter than the present age of the Universe. In this case, we can
assume that all the axions of the level $n$ decayed at the time
$t = \tau_n$ and we find that the photons produced in the decay
have today energy
\be
E_n = \frac{m_n}{2} \left(\frac{3 H_0 \tau_n}{2}\right)^{2/3}
= 0.47 \, \left(\frac{1 \; \mathrm{MeV}}{m_n}\right) \; {\rm keV} \, .
\ee
For example, photons of the X-ray background with energy in the
range $1 - 2$~keV might have be produced by the decay of
cosmological KK axions with mass $250 - 500$~keV. Since the
amount of axions produced via gravitational or misalignment
mechanism are respectively
\be
\frac{\rho_g}{s} &\sim& 0.1 \, M_5^{-21/4} \, , \\
\frac{\rho_m}{s} &\sim& 5 \cdot 10^{29} \Theta_{in}^2 \, M_5^{-21/4} \, .
\ee
Requiring
\be
\frac{B_\gamma \rho_{g,m}}{s}  \lesssim \frac{4\pi}{3}
\frac{1}{s_0}
\int_{1 \; {\rm keV}}^{2 \; {\rm keV}} F_{obs}(E) \, dE \, ,
\label{xray-mis}
\ee
we get the constraint $M_5 \gtrsim 7 \cdot 10^3$~TeV from gravitationally produced axions.

For the misalignment mechanism, if we use equation~(\ref{xray-mis}), we find $M_5 \gtrsim 5 \cdot 10^9 \Theta_{in}^{8/21}$~TeV.  This result however is valid only for small angles, as when $\Theta_{in}$ approaches $\pi$ the mass of the first KK state reaches (and surpasses) 500~keV, which means that today the photons produced in the decay are below 1~keV (see table~\ref{table}).

\subsection{Reionisation}
Photons produced by particles decayed after recombination ($\tau \gtrsim 10^{13}$~s) are redshifted due to the expansion of the Universe and may thus leave the transparency window $E \approx 1 \; {\rm keV} - 1 \; {\rm TeV}$. If this were the case, they would interact with the intergalactic medium and provide an extra source for reionisation. The abundance of the parent particles can be constrained by demanding that the contribution from these photons is still consistent with the optical depth of the last scattering surface obtained from WMAP.

In our case, reionisation can constrain the abundance of KK states with mass around 1~MeV, which decayed at $t \approx 10^{13}$~s into two photons, whose energy today is expected to be about 0.5~keV. Assuming $1/R \ll 1$~MeV, the fraction of energy density of KK axions with mass around 1~MeV is
\be
\Omega_g &\approx& 
1.6 \cdot 10^{13} \, M_5^{-21/4} \, , \nonumber\\
\Omega_m &\approx& 
1.2 \cdot 10^{42} \, \Theta_{in}^2 \, M_5^{-21/4} \, ,
\ee 
respectively from gravitational and misalignment production. Observational data require that~\cite{zhang}\footnote{The parameter $\zeta$ of ref.~\cite{zhang} can be read as $\Omega_{g,m}/\Omega_{dm}$ 
here.}
\be
B_\gamma \Omega_{g,m} \lesssim 10^{-11} \, .
\ee
So, assuming $B_\gamma \approx 1$, we find respectively $M_5 \gtrsim 3 \cdot 10^4$~TeV and $M_5 \gtrsim 1.1 \cdot 10^{10} \Theta_{in}^{8/21}$~TeV.  As already pointed out, the last bound is only good for small initial misalignment, while large angles imply that the first KK state would need to be heavier than about 1~MeV.

\begin{table}
\begin{center}
\begin{tabular}{|l|c|c|c|c|}
\multicolumn{5}{c}{Gravitational production} \\
\multicolumn{5}{c}{} \\
\hline
& Abundance (times $M_5^{21/4}$) & $M_5$ & Highest mass & Mass gap \\
\hline
\hline
Overclosure	& $1.5 \cdot 10^{-6}$	& $17$			& 40 keV	& $10^{-15}$ eV \\
\hline
BBN 1		& $3 \cdot 10^{13}$	& $1.9 \cdot 10^{4}$	& 1 GeV		& 10 $\mu$eV \\
\hline
BBN 2		& $1.7 \cdot 10^{9}$	& $1.0 \cdot 10^{5}$	& 100 MeV	& 1 meV \\
\hline
BBN 3		& $9 \cdot 10^{13}$	& $9 \cdot 10^{4}$	& 1 GeV		& 1 meV \\
\hline
CMB 1		& $4 \cdot 10^{10}$	& $4 \cdot 10^{4}$	& 300 MeV	& 0.1 meV \\
\hline
CMB 2		& $2 \cdot 10^{6}$	& $1.3 \cdot 10^{4}$	& 30 MeV	& 1 $\mu$eV \\
\hline
$N(\nu)$	& $7 \cdot 10^{8}$	& $1.5 \cdot 10^{3}$	& 100 MeV	& 1 neV \\
\hline
$X$-ray		& 0.1			& $7 \cdot 10^{3}$	& 500 keV	& 0.5 $\mu$eV \\
\hline
Reionisation	& 2			& $3 \cdot 10^{4}$	& 1 MeV		& 10 $\mu$eV \\
\hline
\multicolumn{5}{c}{} \\
\multicolumn{5}{c}{Misalignment production} \\
\multicolumn{5}{c}{} \\
\hline
& Abundance (times $M_5^{21/4}$) & $M_5$ (over $\Theta_{in}^{8/21}$) & Highest mass & Mass gap (over $\Theta_{in}^{24/21}$)\\
\hline
\hline
Overclosure	& $1.3 \cdot 10^{27}$	& $3.3 \cdot 10^{7}$	& 40 keV	& 40 keV \\
\hline
BBN 1		& $2 \cdot 10^{37}$	& $7 \cdot 10^{8}$	& 1 GeV		& 300 MeV \\
\hline
BBN 2		& $1.2 \cdot 10^{35}$	& $8 \cdot 10^{9}\,\,\,[5 \cdot 10^8]$	& 100 MeV	& 500 GeV \\
\hline
BBN 3		& $4 \cdot 10^{37}$	& $3 \cdot 10^{9}\,\,\,[1.0 \cdot 10^9]$	& 1 GeV		& 30 GeV \\
\hline
CMB 1		& $7 \cdot 10^{35}$	& $3 \cdot 10^{9}\,\,\,[7 \cdot 10^8]$	& 300 MeV	& 10 GeV \\
\hline
CMB 2		& $4 \cdot 10^{33}$	& $1.9 \cdot 10^{9}\,\,\,[3 \cdot 10^8]$	& 30 MeV	& 10 GeV \\
\hline
$N(\nu)$	& $7 \cdot 10^{34}$	& $1.4 \cdot 10^{8}$	& 100 MeV	& 3 MeV \\
\hline
$X$-ray		& $5 \cdot 10^{29}$	& $5 \cdot 10^{9}\,\,\,[8 \cdot 10^7]$	& 30 GeV	& 100 keV \\
\hline
Reionisation	& $2 \cdot 10^{30}$	& $1.1 \cdot 10^{10}\,\,\,[1.0 \cdot 10^8]$	& 1 TeV		& 500 keV \\
\hline
\multicolumn{5}{c}{} \\
\multicolumn{5}{c}{Supernova} \\
\multicolumn{5}{c}{} \\
\hline
\multicolumn{5}{|c|}{$M_5 \gtrsim 1.5 \cdot 10^{7}$} \\
\hline
\end{tabular}
\end{center}
\caption{Summary of all the constraints analysed in section~\ref{bounds}.  The abundances and limits on $M_5$ are expressed in TeV units.  In square brackets we show the actual lower limit on $M_5$ whenever the minimum mass gap demanded by the constraint is comparable to the highest mass considered in the same band.  Indeed, for large initial $\Theta$, the mass gap is pushed beyond the constraining band, whose upper limit corresponds to the 5D Planck mass shown in square brackets.  The angle for which this happens can be found by comparing the last two columns.}
\label{table}
\end{table}

\subsection{Astrophysics - Supernova bound}
The usual supernova bound arises~\cite{Raffelt:2006cw} from the estimate of the energy loss by the star in the explosion, due to the escape of weakly interacting axions.  The only difference is now that we have $TR$ states (where $T \approx 30$~MeV is the SN temperature) and the 4D limit (fixing $f_4=10^7$~TeV as before) becomes
\be
\frac{f_4}{\sqrt{TR}} \gtrsim 10^5 \; {\rm TeV} \, ,
\ee
that is, $M_5 \gtrsim 1.5 \cdot 10^6$~TeV, valid as long as $TR\gg1$, i.e.\ $M_5 \lesssim 2 \cdot 10^8$~TeV.

\section{Summary and conclusions \label{conclusions}}
In this work we have analysed the cosmology of 5D axions models, in particular in connection with brane world cosmologies.  The main motivation for studying this problem comes from the fact that every time a new bulk field is introduced, the 4D cosmology is automatically affected by the appearance of new KK states;  at the same time the high energy behaviour of the scale factor is known to be different than ordinary FRW cosmology, which has a profund impact on the abundances of particles generated during that epoch.

In addressing this problem we have computed the abundances of KK axions produced by the three main (and most model independent) mechanisms: thermal production, the revitalised gravitational production, and production from misalignment.  It has been shown why we expect that thermal abundances of KK modes is not likely to play any interesting r\^ole, whereas other production mechanisms can put stringent constraints on the parameters of the 5D model, of which we have chosen to single out the 5D Planck mass $M_5$.

Depending on some details of the 5D axion model and its precise (but unknown) cosmologically history around the inflationary epoch (if any), it is possible to constrain $M_5$ to be not below some $10^4$~TeV to $10^8$~TeV, for a fixed $f_4=10^7$~TeV.  This fiducial value for $f_4$ has been chosen since the usual 4D limits are most likely to hold in the 5D setup, especially in 5D, unless some strong fine tuning on the 5D parameters is required.

Although there is no significant improvement as far as the 4D axion decay constant is concerned, we were still able to provide interesting lower limits on $M_5$ (or $f_5$, which is equivalent) that, for instance, preclude the existence of a TeV range (or even $10^3$~TeV) extra dimension if the axion is allowed to propagate in the bulk.  This conclusion relies heavily on the existence of a period of brane expansion in the early Universe, during which the abundances of KK particles can be greatly enhanced.  The only way to completely evade the conditions we have reported is thus to build a model in which the Universe never had equilibrium energy densities high enough to allow for the aforementioned non standard expansion era.  Of course even in this case some production mechanism is still available (gravitational particle production) but it is impossible to draw any conclusion, lacking a precise formulation of such cosmological epoch.  Notice finally that in this paper we have not considered any solitonic production mechanism, which makes our limits conservative.

Astrophysics is also known to be sensitive to the axion properties;  if the axion field is comprised of a tower of different masses, and at least some of them contribute to supernova (or stellar) cooling, then it is possible to constrain $M_5$ in yet another independent way.  Using $f_4=10^7$~TeV we have found that supernov\ae\ limit $M_5$ from below at about $10^7$~TeV, which is, in some cases, the most relevant one.

Before concluding, a note on the r\^ole of $f_4$ in this analysis.  We have fixed this parameter before discussing the cosmological history of the KK axions, but it is easy to see what would have happened, had we picked up a different value for it.  Consider first a larger $f_4$.  This means that axion interactions are weaker, and that the same KK state would decay later, rendering its contribution potentially more worrisome.  This is indeed the case, and one can see that tighter limits would be obtained in this case (although if we confine ourselves to the 4D allowed range~\ref{bound-f} the differences are mild).  Had we chosen a much smaller $f_4$ instead the constraints coming from gravitational and misalignment productions would be relaxed.  However, in that case thermal equilibrium would come back in, as the window at which the number of equilibrium degrees of freedom is dominated by $n_{KK}$ broadens to much smaller $M_5$, which means that (despite the faster decay rate) the enormous number of KK states reaching equilibration would survive well after 1s, and the 5D Planck mass would need to be again pushed up enough for the first state to decay before BBN.

\begin{acknowledgments}
One of us (FU) wishes to acknowledge the Institute for the Physics and Mathematics of the Universe, University of Tokyo, for support and hospitality while this work was being completed.  CB is supported by World Premier International Research Center Initiative (WPI Initiative), MEXT, Japan.
\end{acknowledgments}

\end{document}